\documentclass[twocolumn,aps,prl,superscriptaddress,amsmath,amssymb]{revtex4-2}
\usepackage{amsfonts}
\usepackage{bbm}
\usepackage{graphicx}
\usepackage{dcolumn}
\usepackage{bm}
\usepackage{epstopdf}
\usepackage{color}

\begin{document}

\title{Transverse spin selectivity in helical nanofibers prepared without any chiral molecule}
\author{Chenchen Wang}
\thanks{These authors contributed equally to this work.}
\affiliation{CAS Key Laboratory of Nanosystem and Hierarchical Fabrication, CAS Center for Excellence in Nanoscience, National Center for Nanoscience and Technology, Beijing 100190, China}
\affiliation{University of Chinese Academy of Sciences, Beijing 100049, China}

\author{Zeng-Ren Liang}
\thanks{These authors contributed equally to this work.}
\affiliation{Hunan Key Laboratory for Super-microstructure and Ultrafast Process, School of Physics, Central South University, Changsha 410083, China}

\author{Xiao-Feng Chen}
\affiliation{Hunan Key Laboratory for Super-microstructure and Ultrafast Process, School of Physics, Central South University, Changsha 410083, China}

\author{Ai-Min Guo}
\email{aimin.guo@csu.edu.cn}
\affiliation{Hunan Key Laboratory for Super-microstructure and Ultrafast Process, School of Physics, Central South University, Changsha 410083, China}

\author{Guanghao Ji}
\affiliation{CAS Key Laboratory of Nanosystem and Hierarchical Fabrication, CAS Center for Excellence in Nanoscience, National Center for Nanoscience and Technology, Beijing 100190, China}
\affiliation{University of Chinese Academy of Sciences, Beijing 100049, China}

\author{Qing-Feng Sun}
\affiliation{International Center for Quantum Materials, School of Physics, Peking University, Beijing 100871, China}
\affiliation{Hefei National Laboratory, Hefei 230088, China}

\author{Yong Yan}
\email{yany@nanoctr.cn}
\affiliation{CAS Key Laboratory of Nanosystem and Hierarchical Fabrication, CAS Center for Excellence in Nanoscience, National Center for Nanoscience and Technology, Beijing 100190, China}
\affiliation{University of Chinese Academy of Sciences, Beijing 100049, China}
\affiliation{Department of Chemistry, School of Chemistry and Biological Engineering, University of Science and Technology Beijing, Beijing 100083, China}

\date{\today}

\begin{abstract}
In the last decade, chirality-induced spin selectivity (CISS) has been attracting extensive interest. However, there still exists a large gap between experiments and quantitative theoretical results, and the microscopic mechanism of CISS, especially transverse CISS where electrons are injected perpendicular to the helix axis of chiral molecules, remains elusive. Here, we address these issues by performing a combined experimental and theoretical study on conducting polyaniline helical nanofibers which are synthesized in the absence of any chiral species. Large spin polarization is measured in both left- and right-handed nanofibers for electrons injected perpendicular to their helix axis, which is comparable to the value of parallel electron injection in other chiral molecules, and it will be reversed by switching the handedness between two enantiomers. We develop a theoretical model with extremely weak spin-orbit coupling arising exclusively from electron propagation between neighboring polyanilines and the numerical results are quantitatively consistent with the experimental data. Our results demonstrate that the supramolecular handedness is sufficient for spin-selective electron transmission in chiral molecules assembled from achiral monomers and the mechanism of transverse CISS is revealed.

\end{abstract}

\maketitle

{\it Introduction.}---Since the original experiment of asymmetric scattering of spin-polarized photoelectrons by chiral molecules \cite{rk1}, chirality-induced spin selectivity (CISS) has been receiving extensive attention \cite{nr1,nr2,nr3} and showed great promise in spintronics \cite{bdo1,kg1,abh1,abh2,gn1}, enantioseparation \cite{bgk1,tf1,dia1,sk1,wc1,frj1} and understanding biological processes \cite{nr4,osf1,bbp1}. This CISS refers to longitudinal CISS, where large spin polarization is detected for unpolarized electrons injected parallel to the helix axis of chiral molecules, and it was demonstrated in various chiral species, including double-stranded DNA \cite{gb1,xz1,ztj1,ajm1,ms1,dtk1}, single-helical protein/peptide \cite{md1,kem1,eh1,aac1,ms2,lt1}, supramolecular polymers \cite{mpc1,jl1,mis1,kc1}, and other organic compounds \cite{kv1,km1,smr1,msp1,mpc2,bbp2,bbp3,cc1,sm1,lhp1,kyh1,qq1,nar1,mpv1,kch1}. Many theories were proposed to elucidate this longitudinal CISS \cite{gam1,me1,eaa1,gam2,gam3,vs1,mas1,mvv1,mk1,ds1,gm1,dgf1,uy1,zl1,yx1,hpj1,frj2,as1,ly1,woy1}.

Very recently, Naaman {\it et al.} measured the spin transport through supramolecular polymers containing achiral monomers, finding that they present spin filtering for electrons injected perpendicular to the helix axis of these polymers \cite{mak1}. This is named as transverse CISS and experimentally confirmed in another supramolecular structure \cite{sy1}. In most previous experiments, however, chiral species were involved in preparing left- and right-handed (supra)molecules \cite{gb1,xz1,ztj1,ajm1,ms1,dtk1,md1,kem1,eh1,aac1,ms2,lt1,mpc1,mis1,kc1,kv1,km1,smr1,msp1,mpc2,bbp2,bbp3,cc1,sm1,lhp1,kyh1,qq1,nar1,jl1,mpv1,kch1,mak1}, and the residual ones may somewhat affect their spin transport properties. In particular, all previous theoretical studies focused on the longitudinal CISS of chiral molecules \cite{gam1,me1,eaa1,gam2,gam3,vs1,mas1,mvv1,mk1,ds1,gm1,dgf1,uy1,zl1,yx1,hpj1,frj2,as1,ly1,woy1}, and their transverse CISS has not yet been addressed. And a significant gap remains between experimental observations and quantitative theoretical simulations \cite{wdh1,ef1,acd1}.

\begin{figure}
\includegraphics[width=0.48\textwidth]{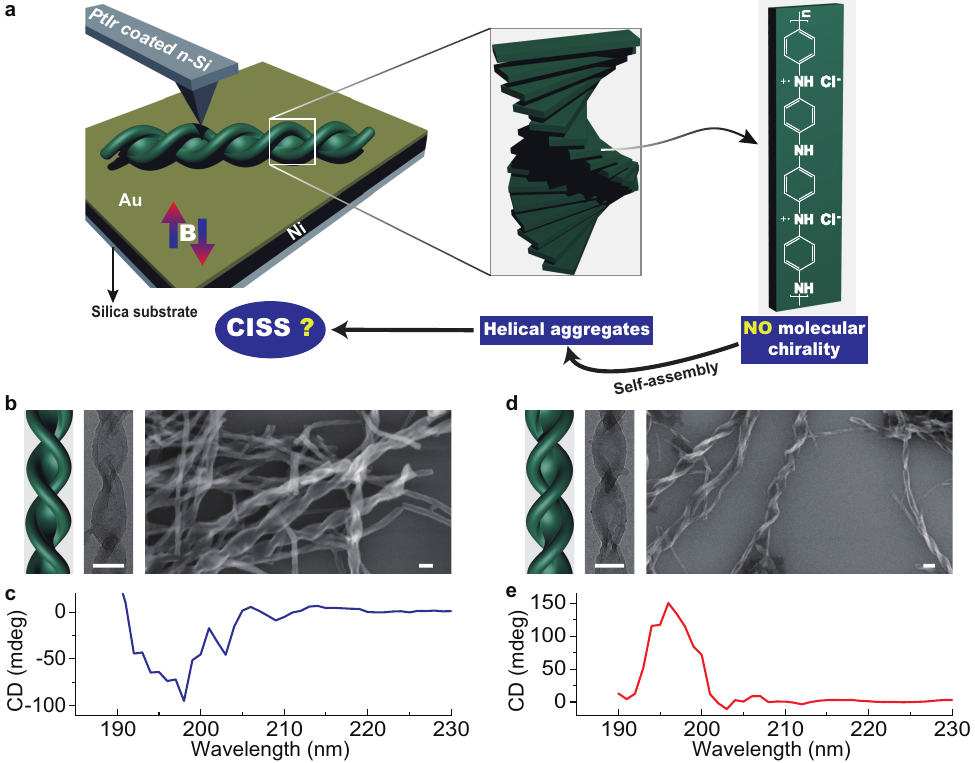}
\caption{\label{fig1} Experimental setup, materials, and CD spectra. (a) Schematics of mc-AFM measurements on PANI left-handed nanofiber (left) and molecular structure of HCl doped PANI (right). 
Schemes (left), TEM (middle, scale bars, 50 nm), and SEM (right, scale bars, 100 nm) images of (b) left- and (d) right-handed nanofibers. (c) and (e) CD spectra of both nanofibers.}
\end{figure}

In this Letter, we study the transverse CISS in helical nanofibers prepared from achiral polyanilines (PANIs) without any chiral species. 
By performing magnetic-conductive atomic force microscopy (mc-AFM), the spin-dependent electron transport is measured by injecting electrons normal to the nanofiber helix axis (Fig.~\ref{fig1}a), finding that the spin polarization achieves $\sim$70\% and will be reversed by switching the nanofiber handedness. 
A theoretical model with extremely weak spin-orbit coupling (SOC) is firstly proposed to elucidate this transverse CISS, where the SOC only exists for electron hopping between two neighboring PANIs and vanishes for electron propagation within a single PANI. The theoretical results are quantitatively consistent with the experiment, filling the gap between experiments and theory.

{\it Experiment.}---The PANI helical nanofibers are synthesized according to the previous method \cite{zcq1,zcq2}, where aniline monomers are oxidized by ammonium persulfate in the mixed solvent of water and isopropyl alcohol \cite{sm}. A small amount of hydrochloric acid is added as the dopant. In the whole polymerization process, there are no chiral reagents and solvents. Besides, the reaction container is highly symmetric (cylinder glass vial) which will not impose any structural influence on the polymer growth. Moreover, the polymerization has proceeded in a homogeneous solution without agitation, and thus the interfacial effect and vortex flow commonly used to assemble achiral materials/molecules into chiral aggregates are not applicable.

Interestingly, by adjusting the solvent ratio of isopropyl alcohol and water, enantiomeric excess of PANI helical nanofibers is produced. For example, with 35\% isopropyl alcohol (v/v) in the mixture solvent, almost pure left-handed nanofibers are obtained (Figs.~\ref{fig1}b and~\ref{fig1}c). Increasing alcohol content to 45\% changes the polarity environment, and subsequently the nanofiber handedness becomes inverse (Figs.~\ref{fig1}d and~\ref{fig1}e). The symmetry breaking and chirality inversion are probably due to the cooperative effect of hydrogen bonding and $\pi$-stacking between neighboring PANIs \cite{zcq1,zcq2}.

\begin{figure}
\includegraphics[width=0.48\textwidth]{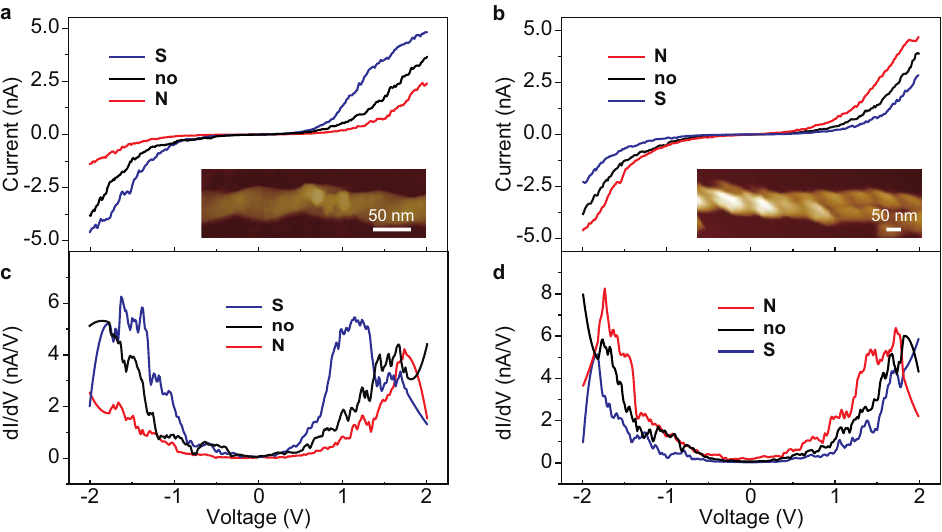}
\caption{\label{fig2} Transverse CISS of PANI helical nanofibers. Current-voltage characteristics of (a) left- and (b) right-handed nanofibers when the bottom Au/Ni electrode is magnetized by a magnet with its south pole (blue curves) or north pole (red curves) pointing upward. The measurement is also conducted in the absence of magnetization (black curves). The insets in (a) and (b) are the AFM images of these nanofibers with scale bars being 50 nm. All the I-V curves are averaged over 50 measurements. Corresponding $dI/dV \sim V$ for (c) left- and (d) right-handed nanofibers.}
\end{figure}

Then, the spin transport through these nanofibers is investigated by means of mc-AFM when electrons are injected normal to their helix axis. The bottom electrode is ferromagnetic, which is composed of a layer of 150 nm-thick Ni and a layer of 10 nm-thick Au on the top (Fig.~\ref{fig1}a). With the magnetic field, electrons in the Ni layer are polarized and injected into the PANI nanofibers with an excess of one type of spin. The typical current-voltage characteristics of PANI left- and right-handed nanofibers are shown in Figs.~\ref{fig2}a and~\ref{fig2}b, respectively. For the left-handed nanofiber, a higher current is recorded when the bottom Au/Ni electrode is magnetized by a magnet (0.35 T for at least 30 mins) with its south pole pointing upward (see the blue curve of Fig.~\ref{fig2}a). 
However, when the bottom Ni electrode is polarized with the reversed magnetic field direction, the current is decreased (see the red curve of Fig.~\ref{fig2}a). Before the magnetization of the Ni electrode, the current is measured for reference (see the black curve of Fig.~\ref{fig2}a).

Remarkably, these observations in the left-handed nanofibers are reversed in the right-handed ones, where the north (south) pole upward magnetization refers to higher (lower) current (Fig.~\ref{fig2}b). This apparent spin-selective transmission is confirmed by plotting $dI/dV \sim V$ which can qualitatively describe the effective ``barrier'' of electron hopping within these $\pi$-conjugated systems. For the left-handed nanofiber, the lowest ``barrier'' is found when the bottom ferromagnetic electrode is south-pole-upward magnetized (see the blue curve of Fig.~\ref{fig2}c). However, this direction magnetization observes the highest “barrier” in the right-handed nanofiber device (see the blue curve of Fig.~\ref{fig2}d). In both helical nanofibers, the ``barrier'' heights of non-magnetized devices lie between two magnetization cases (see the black curves of Figs.~\ref{fig2}c and~\ref{fig2}d). 

\begin{figure}
\includegraphics[width=0.48\textwidth]{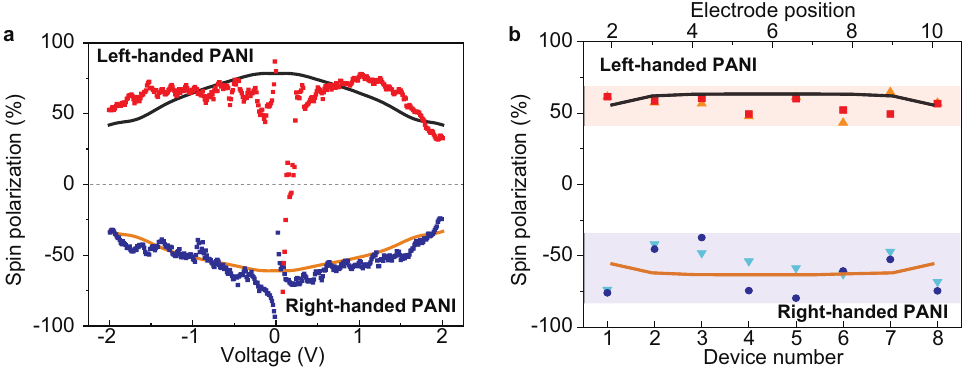}
\caption{\label{fig3} Spin selectivity of PANI helical nanofibers. (a) Spin polarization versus bias voltage for PANI left- (red marker for experimental data and black line for theoretical results) and right-handed (blue marker for experimental data and orange line for theoretical results) nanofibers. (b) Experimental (markers, bottom x-axis) and theoretical (lines, top x-axis) spin polarization for different nanofibers and/or different coupling locations of the upper electrode  in the same nanofiber.
}
\end{figure}

The spin selective capability of the PANI nanofibers is then evaluated by calculating the current ratio $SP=(I_{\rm S}-I_{\rm N})/(I_{\rm S}+I_{\rm N})$, where $I_{\rm S}$ and $I_{\rm N}$ are the measured currents when the bottom Ni electrode is magnetized with the south and north poles of the magnet pointing upward respectively. Positive and negative spin polarizations are obtained for PANI left- and right-handed nanofibers (see the red and blue markers of Fig.~\ref{fig3}a), respectively. The highest SPs for both nanofibers are approximately 70\%, which is comparable to the value for longitudinal CISS observed in other devices fabricated by chiral building blocks \cite{mpc1,mis1,kc1}. Besides, the SP gradually declines with increasing the bias voltage.

To confirm the reliability of our results, all of the current-voltage curves are averaged over independent 50 measurements, as shown in the Supplementary Material \cite{sm}. Importantly, this measurement is repeated for additional seven times by a random selection of different helical nanofibers and/or different locations of the same nanofiber, namely, eight independent tests for both left- and right-handed nanofibers. Similar current-voltage characteristics are observed and all of the devices show spin-dependent transmission behaviors \cite{sm}. The spin polarization of all devices is summarized as the markers in Fig.~\ref{fig3}b. All of the left- and right-handed nanofibers display, respectively, positive and negative SP values with relatively small deviations.

\begin{figure}
\includegraphics[width=0.48\textwidth]{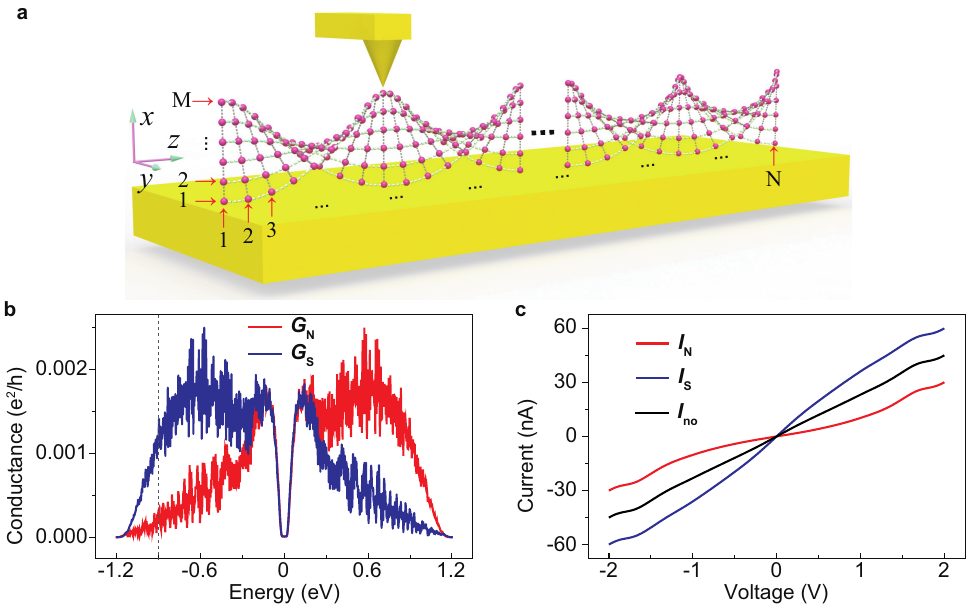}
\caption{\label{fig4} Schematic of the theoretical model of the left-handed nanofiber and its numerical results. (a) Schematic of the left-handed nanofiber contacted by a single upper non-magnetic electrode and several lower magnetic electrodes, where each ball denotes an aniline and self-assembles into an achiral PANI (straight line parallel to the x-y plane). (b) Energy-dependent conductance $G_{\rm N}$ with upward magnetization and $G_{\rm S}$ with downward magnetization. (c) Current-voltage characteristics with upward magnetization (red line), downward magnetization (blue line), and no magnetization (black line). The upward direction is parallel to the x axis.}
\end{figure}

{\it Model.}---A theoretical model with extremely weak SOC is then put forward to unveil the transverse CISS in PANI helical nanofibers, the schematic of which is shown in Fig.~\ref{fig4}a. As there are no magnetic atoms in PANIs, the SOC will be the key factor for the CISS \cite{gam1,gam2,gam3}. Specifically, we assume that the SOC exists only in the chiral coupling between neighboring PANIs but it vanishes inside an achiral PANI. In Fig.~\ref{fig4}a, each straight line parallel to the x-y plane denotes an achiral PANI, and all the lines self-assemble into a chiral nanofiber whose helix axis points along the z axis. We consider a nanofiber device contacted by a single upper non-magnetic electrode and several lower magnetic electrodes, and its electron transport can be described by the Hamiltonian
\begin{eqnarray}\mathcal{H} = \mathcal{H}_{\rm P} +\mathcal{H}_{\rm SOC} +\mathcal{H}_{\rm U}+ \mathcal{H}_{\rm L}+\mathcal{H}_{\rm d}.
\end{eqnarray}
$\mathcal{H}_{\rm P}$ is the tight-binding Hamiltonian of the nanofiber and reads
\begin{eqnarray}
\mathcal{H}_{\rm P} =& \sum_{n,m} (\varepsilon_{nm} c_{nm}^{\dag} c_{nm}+ t_{\perp} c_{nm}^{\dag} c_{nm+1} +t_{\parallel  } c_{nm}^{\dag} c_{n+1m}+{\rm H.c.}),\label{eq2}
\end{eqnarray}
where $c_{nm}^{\dag}=(c_{nm\uparrow }^{\dag}, c_{nm\downarrow }^{\dag})$ is the creation operator at site $(n,m)$, with $n$ the PANI index and $m$ the site index of a PANI (Fig.~\ref{fig4}a). The number of PANIs is $N$, and that of aniline in a PANI is $M$. $\varepsilon_{nm}$ is the on-site energy, $t_{\perp}$ the nearest-neighbor hopping integral in a PANI, and $t_{\parallel}$ the one between two successive PANIs. $\mathcal{H}_{\rm SOC}$ is the SOC Hamiltonian and reads
\begin{eqnarray}
\begin{aligned}
\mathcal{H}_{\rm SOC} =& \sum_{n=1}^{N-1} \sum_{m=1}^{M}i s c_{nm}^{\dag} \frac{r_m}{r_0} \{[\sigma_{x} (\sin\varphi_{nm} +\sin\varphi_{n+1m}) \\& + \sigma_{y}(\cos\varphi_{nm} +\cos\varphi_{n+1m}) ]\sin\theta_{m}\\&+  2\sigma_{z} \sin\beta\cos\theta_{m}\} c_{n+1m}+ {\rm H.c.}\label{eq3}
\end{aligned}
\end{eqnarray}
Here, $s$ is the SOC strength, $r_m$ the distance from site $(n,m)$ to the helix axis (z axis), $r_0$ the lattice constant of a PANI, $\varphi _{nm} =(n-1)\Delta \varphi +[3-\mathrm{sgn}(M+1-2m)  ]\pi/2$ the azimuth angle of site $(n,m)$, $\theta _m =\arcsin[\Delta h/l_m ] $ the space angle between the vector from site $(n,m)$ to $(n+1,m)$ and the x-y plane, $\beta  =[\mathrm{sgn}(\Delta \varphi )\pi - \Delta \varphi]/2$ the deviation angle \cite{sm}, and $\sigma _{x,y,z}$ the Pauli matrices. $\Delta \varphi$ and $\Delta h$ are, respectively, the twist angle and the stacking distance between two successive PANIs, $l_m=\sqrt{[(M+1-2m)a\sin (\Delta \varphi /2)]^2+(\Delta h)^2} $ the distance from site $(n,m)$ to $(n+1,m)$ \cite{sm}, and sgn the sign function. This SOC originates from electron propagation under chiral potentials \cite{gam1,gam2}. Specifically, the SOC Hamiltonian arises only when electrons transport between two successive PANIs because of the nanofiber chirality, whereas it vanishes for electron propagation in a single, achiral PANI. The third and fourth terms denote, respectively, the single upper non-magnetic electrode ($\mathcal{H}_ {\rm U}$) and several lower magnetic electrodes ($\mathcal{H}_{\rm L}$) and their couplings to the nanofiber, and are expressed as $\mathcal{H}_{\rm U} = \sum_{k}\varepsilon_{{\rm U} k}a_{{\rm U} k}^{\dag} a_{{\rm U} k}+\tau _{{\rm U} } a_{{\rm U} k}^{\dag} c_{n_{\rm u} m_{\rm u}}+{\rm H.c.} , \mathcal{H}_ {\rm L} = \sum_{j, k}\varepsilon_{{\rm L}_j k}a_{{\rm L}_j k}^{\dag} a_{{\rm L}_j k}+\tau _{{\rm L}_j }a_{{\rm L}_j k}^{\dag} c_{n_j m_j}+{\rm H.c.}.$ Here, $\tau _{\rm U}$ $(\tau _{{\rm L}_j })$ is the coupling between the nanofiber and the upper non-magnetic ($j$th lower magnetic) electrode. We point out that the upper electrode is connected to a single, topmost site, and the lower electrodes are coupled to all of the bottommost sites. The electrons flowing between the upper and lower electrodes are injected perpendicular to the nanofiber helix axis. Finally, the last term simulates the electron leakage from the nanofiber to the environment and is written as $\mathcal{H}_{\rm d} = \sum_{n,m,k} (\varepsilon_{nmk} d_{nmk}^{\dag} d_{nmk}+ t _{\rm d}d_{nmk}^{\dag} c_{n m}+{\rm H.c.})$.

By combining the Landauer-B\"{u}ttiker formula and the Green's function method \cite{das1,ynx1}, the Hamiltonian $\mathcal{H}$ of this nanofiber device can be solved, and both conductance $G_{\rm N}$ and current $I_{\rm N}$ can be obtained when the lower electrodes are magnetized upward (parallel to the x axis), so do $G_{\rm S}$ and $I_{\rm S}$ for downward magnetization \cite{sm}. In the numerical calculations, the structural parameters are consistent with the experiments \cite{zcq1,zcq2} and the model parameters are taken from previous works \cite{sm,tof1,cs1}. We point out that the helical nanofibers exhibit pronounced CISS effect even for extremely weak SOC of 0.6 meV.

{\it Numerical Results.}---Figure~\ref{fig4}b shows $G_{\rm N}$ and $G_{\rm S}$ for the PANI left-handed nanofiber contacted by the upper electrode at the second topmost site. Here, the conductance is obtained at zero bias voltage. Although there is no SOC for electron propagation in an achiral PANI, $G_{\rm N}$ is different from $G_{\rm S}$ over almost the whole energy spectrum, indicating the CISS effect of the chiral nanofiber. This is consistent with the experiment, demonstrating that the chirality of the nanofiber instead of the constituent of PANI plays an important role in generating the CISS effect. Besides, the conductances satisfy $G_{\rm N}(E)=G_{\rm S}(-E)$, owing to the electron-hole-type symmetry \cite{gam4}.

We then turn to calculate the current by choosing the Fermi energy as the fitting parameter. Figure~\ref{fig4}c displays the current $I_{\rm N}$ and $I_{\rm S}$ for the left-handed nanofiber by setting the Fermi energy $E_{\rm F}=-0.9$ eV (see the dashed line of Fig.~\ref{fig4}b), where the current without magnetization is shown for reference. It is clear that $I_{\rm S}$ is always greater than $I_{\rm N}$, indicating that the spin-down electrons can propagate through the left-handed nanofiber more efficiently than the spin-up ones. In the absence of magnetization, the current falls between $I_{\rm S}$ and $I_{\rm N}$. All these results are consistent with the experiment. Notice that the calculated current is one order of magnitude larger than the experimental data, because the disorder effect is ignored.

We also calculate the SP by changing the bias voltage $V_{\rm b}$ (see the black and orange lines of Fig.~\ref{fig3}a). Here, the upper electrode is contacted at the second topmost site for the left-handed nanofiber and at the middle one for the right-handed nanofiber \cite{sm}. It is clear that the numerical results are quantitatively consistent with the experimental data (Fig.~\ref{fig3}a). Specifically, the SP is symmetric with respect to the zero bias voltage, and it reaches the maximum of 69.6\% (-73.1\%) for the left- (right-) handed nanofiber. This high SP arises from the quantum interference among different transport pathways and the accumulation of spin selectivity for electron propagation in such large systems \cite{hpj1,ptr}. Besides, the SP decreases with $V_{\rm b}$ for both nanofibers, which can be understood as follows. We take the left-handed nanofiber as an example. When the bias voltage approaches zero, the electron transport is determined by the electronic states around the Fermi level and the left-handed nanofiber exhibits high SP because of large difference between $G_{\rm N}$ and $G_{\rm S}$ around the Fermi energy (Fig.~\ref{fig4}b). By increasing $V_{\rm b}$, the electronic states within the range of $[E_{\rm F}-eV_{\rm b}/2, E_{\rm F}+eV_{\rm b}/2]$ will contribute to the current. Nevertheless, the difference between $G_{\rm N}$ and $G_{\rm S}$ shrinks when the electron energy $E$ deviates from $E_{\rm F}$, and $G_{\rm N}$ can even be greater than $G_{\rm S}$ for $E> 0.2$ eV (Fig.~\ref{fig4}b). As a result, the SP of the left-handed nanofiber decreases with increasing $V_{\rm b}$, so does the SP of the right-handed nanofiber.

We point out that, although the SP is obtained along the x axis instead of the helix axis (z axis), it is exactly opposite for the left- and right-handed nanofibers when the upper electrode is connected to the same site of both nanofibers \cite{sm}. This extends previous studies on various chiral molecules that the SP measured along the helix axis will be reversed for two enantiomers \cite{kv1,ztj1,km1,sm1,gam1,gam2}. Finally, the SP is computed by coupling the upper electrode to different topmost sites of both nanofibers at $V_{\rm b}$ = 1 V (see the black and orange lines of Fig.~\ref{fig3}b). One can see that the absolute value of the SP of both nanofibers ranges from 55.7\% to 63.4\% by changing the position of the upper electrode, quantitatively consistent with the experiment. Besides, the SP for both nanofibers is symmetric with respect to the middle topmost site due to the structural symmetry.

In summary, we have demonstrated that the monomer chirality is not necessary for spin-selective electron transmission in chiral molecules. Large spin selectivity is observed in conducting polyaniline helical nanofibers prepared without the involvement of any chiral species when electrons are injected normal to the helix axis. The numerical calculations support the experiment, showing that the theoretical model, where spin-orbit coupling vanishes inside a single achiral polyaniline, can quantitatively explain the CISS effect for transverse spin injection.

We gratefully acknowledge support from the National Natural Science Foundation of China (52273169, 12274466, 11874428, 11921005, and 12374034), the National Key Research and Development Program of China (2021YFA1200302), the Strategic Priority Research Program of the Chinese Academy of Sciences (XDB36000000), the Innovation Program for Quantum Science and Technology (2021ZD0302403), the Hunan Provincial Science Fund for Distinguished Young Scholars (2023JJ10058), and the High Performance Computing Center of Central South University.

\end{document}